\documentclass[11pt,a4paper]{article}
\usepackage[hyperref]{emnlp2020}
\usepackage{times}
\usepackage{latexsym}

\usepackage{microtype}
\usepackage{amssymb}
\usepackage{amsmath}
\usepackage{blindtext}
\usepackage{booktabs}
\usepackage{multirow}
\usepackage{graphicx}
\usepackage{subcaption}
\usepackage{algorithm}
\usepackage{algorithmic}
\usepackage{enumitem}

\aclfinalcopy 
\def\aclpaperid{349} 

\newcommand\run[1]{\texttt{\small #1}}

\usepackage{array}
\newcolumntype{H}{>{\setbox0=\hbox\bgroup}c<{\egroup}@{}}

\title{Covidex: Neural Ranking Models and Keyword Search Infrastructure\\ for the COVID-19 Open Research Dataset}

\author{Edwin Zhang,$^{1}$ Nikhil Gupta,$^{1}$ Raphael Tang,$^{1}$ Xiao Han,$^{1}$ Ronak Pradeep,$^{1}$ Kuang Lu,$^{2}$  \\
{\bf Yue Zhang},$^{2}$ {\bf Rodrigo Nogueira,}$^{1}$ {\bf Kyunghyun Cho,}$^{3,4}$ {\bf Hui Fang},$^{2}$ \and {\bf Jimmy Lin}$^1$\\[0.2cm]
$^1$ University of Waterloo \quad $^2$ University of Delaware \\
$^3$ New York University \quad $^4$ CIFAR Associate Fellow
}

\date{}

\begin{document}
\maketitle

\begin{abstract}
We present Covidex, a search engine that exploits the latest neural ranking models to provide information access to the COVID-19 Open Research Dataset curated by the Allen Institute for AI.
Our system has been online and serving users since late March 2020.
The Covidex is the user application component of our three-pronged strategy to develop technologies for helping domain experts tackle the ongoing global pandemic.
In addition, we provide robust and easy-to-use keyword search infrastructure that exploits mature fusion-based methods as well as standalone neural ranking models that can be incorporated into other applications.
These techniques have been evaluated in the ongoing TREC-COVID challenge:\ Our infrastructure and baselines have been adopted by many participants, including some of the highest-scoring runs in rounds 1, 2, and~3.
In round 3, we report the highest-scoring run that takes advantage of previous training data and the second-highest fully automatic run.
\end{abstract}

\section{Introduction}

As a response to the worldwide COVID-19 pandemic, on March 13, 2020, the Allen Institute for AI (AI2) released the COVID-19 Open Research Dataset (CORD-19).\footnote{\url{www.semanticscholar.org/cord19}}
With regular updates since the initial release (first weekly, then daily), the corpus contains around 188,000 scientific articles (as of July 12, 2020), including most with full text, about COVID-19 and coronavirus-related research more broadly (for example, SARS and MERS).
These articles are gathered from a variety of sources, including PubMed, a curated list of articles from the WHO, as well as preprints from arXiv, bioRxiv, and medRxiv.
The goal of the effort is ``to mobilize researchers to apply recent advances in natural language processing to generate new insights in support of the fight against this infectious disease.''
We responded to this call to arms.

As motivation, we believe that information access capabilities (search, question answering, etc.) can be applied to provide users with high-quality information from the scientific literature, to inform evidence-based decision making and to support insight generation.
Examples include public health officials assessing the efficacy of wearing face masks, clinicians conducting meta-analyses to update care guidelines based on emerging studies, and virologist probing the genetic structure of COVID-19 in search of vaccines.
We hope to contribute to these efforts via a three-pronged strategy:

\begin{enumerate}[leftmargin=*]
\itemsep0em 

\item Despite significant advances in the application of neural architectures to text ranking, keyword search (e.g., with ``bag of words'' queries) remains an important core technology.
Building on top of our Anserini IR toolkit~\cite{Yang_etal_JDIQ2018}, we have released robust and easy-to-use open-source keyword search infrastructure that the broader community can build on.

\item Leveraging our own infrastructure, we explored the use of sequence-to-sequence transformer models for text ranking, combined with a simple classification-based feedback approach to exploit existing relevance judgments.
We have also open sourced all these models, which can be integrated into other systems.

\item Finally, we package the previous two components into Covidex, an end-to-end search engine and browsing interface deployed at \url{covidex.ai}, initially described in~\citet{Zhang_etal_arXiv2020_Covidex}.

\end{enumerate}

\noindent
All three efforts have been successful.
In the ongoing TREC-COVID challenge, our infrastructure and baselines have been adopted by many teams, which in some cases have submitted runs that scored higher than our own submissions.
This illustrates the success of our infrastructure-building efforts (1).
In the latest round 3 results, we report the highest-scoring run that exploits relevance judgments in a user feedback setting and the second-highest fully automatic run, affirming the quality of our own ranking models (2).
Finally, usage statistics offer some evidence for the success of our deployed Covidex search engine (3).

\section{Ranking Components}

Multi-stage search architectures represent the most common design for modern search engines, with work in academia dating back over a decade~\cite{Matveeva_etal_SIGIR2006,Wang_etal_SIGIR2011,Asadi_Lin_SIGIR2013}.
Known production deployments of this design include the Bing web search engine~\cite{Pedersen_SIGIR2010} as well as Alibaba's e-commerce search engine~\cite{LiuShichen_etal_SIGKDD2017}.

The idea behind multi-stage ranking is straightforward:\ instead of a monolithic ranker, ranking is decomposed into a series of stages.
Typically, the pipeline begins with an initial retrieval stage, most often using bag-of-words queries against an inverted index.
One or more subsequent stages reranks and refines the candidate set successively until the final results are presented to the user.
The multi-stage design provides a clean interface between keyword search, neural reranking models, and the user application.

This section details individual components in our architecture.
We describe later how these building blocks are assembled in the deployed system (Section~\ref{section:covidex}) and for TREC-COVID (Section~\ref{section:results}).

\subsection{Keyword Search}
\label{section:keyword}

In our design, initial retrieval is performed by the Anserini IR toolkit~\cite{Yang_etal_SIGIR2017,Yang_etal_JDIQ2018},\footnote{\url{anserini.io}} which we have been developing for several years and powers a number of our previous systems that incorporate various neural architectures~\cite{Yang_etal_NAACL2019demo,Yilmaz_etal_EMNLP2019}.
Anserini represents an effort to better align real-world search applications with academic information retrieval research:\ under the covers, it builds on the popular and widely-deployed open-source Lucene search library, on top of which we provide a number of missing features for conducting research on modern IR test collections.

Anserini provides an abstraction for document collections, and comes with a variety of adaptors for different corpora and formats:\ web pages in WARC containers, XML documents in tarballs, JSON objects in text files, etc.
Providing keyword search capabilities over CORD-19 required only writing an adaptor for the corpus that allows Anserini to ingest the documents.

An issue that immediately arose with CORD-19 concerns the granularity of indexing, i.e., what should we consider to be a ``document'' as the ``atomic unit'' of indexing and retrieval?
One complication is that the corpus contains a mix of articles that vary widely in length, not only in terms of natural variations (scientific articles of varying lengths, book chapters, etc.), but also because the full text is not available for some articles.
It is well known in the IR literature, dating back several decades (e.g.,~\citealt{Singhal96}), that length normalization plays an important role in retrieval effectiveness.

Guided by previous work on searching full-text articles~\cite{Lin_BMCBioinformatics2009}, we explored three separate indexing schemes:

\begin{itemize}[leftmargin=*]
\itemsep0em 

\item An index comprised of only titles and abstracts.

\item An index comprised of each full-text article as a single, individual document; articles without full text contained only titles and abstracts.

\item A paragraph-level index structured as follows:\ each full-text article is segmented into paragraphs and for {\it each} paragraph, we created a ``document'' comprising the title, abstract, and that paragraph.
The title and abstract alone comprised an additional ``document''.
Thus, a full-text article with $n$ paragraphs yields $n+1$ separate retrieval units in the index.

\end{itemize}

\noindent To be consistent with standard IR parlance, we call each of these retrieval units a document, in a generic sense, despite their composite structure.
Following best practice, documents are ranked using BM25~\cite{Robertson94}.
The relative effectiveness of each indexing scheme, however, is an empirical question.

With the paragraph index, a query is likely to retrieve multiple paragraphs from the same underlying article; since the final task is to rank articles, we take the highest-scoring paragraph across all retrieved results to produce a final ranking.
Furthermore, we can combine these multiple representations to capture different ranking signals using fusion techniques, which further improves effectiveness; see Section~\ref{section:results} for details.

Since Anserini is built on top of Lucene, which is implemented in Java, it is designed to run on the Java Virtual Machine (JVM).
However, Tensor\-Flow~\cite{abadi2016tensorflow} and PyTorch~\cite{paszke2019pytorch}, 
the two most popular neural network toolkits today, use Python as their main language.
More broadly, with its diverse and mature ecosystem, Python has emerged as the language of choice for most data scientists today.
Anticipating this gap, we have been working on Pyserini,\footnote{\url{pyserini.io}} Python bindings for Anserini, since late 2019~\cite{Yilmaz_etal_SIGIR2020}.
Pyserini is released as a well-documented, easy-to-use Python module distributed via PyPI and easily installable via \texttt{pip}.\footnote{\url{pypi.org/project/pyserini/}}

Putting everything together, we provide the community keyword search infrastructure by sharing code, indexes, as well as baseline runs.
First, all our code is available open source.
Second, we share regularly updated pre-built versions of CORD-19 indexes, so that users can replicate our results with minimal effort.
Finally, we provide baseline runs for TREC-COVID that can be directly incorporated into other participants' submissions.

\subsection{Rerankers}
\label{section:reranker}

In our infrastructure, the output of Pyserini is fed to rerankers that aim to improve ranking quality.
We describe three different approaches:\ two are based on neural architectures, and the third exploits relevance judgments in a feedback setting using a classification approach.

\smallskip \noindent {\bf monoT5.}
Despite the success of BERT for document ranking~\cite{Dai_Callan_SIGIR2019,MacAvaney_etal_SIGIR2019,Yilmaz_etal_EMNLP2019}, there is evidence that ranking with sequence-to-sequence models can achieve even better effectiveness, particularly in zero-shot and other settings with limited training data~\cite{Nogueira_etal_arXiv2020_T5}, such as for TREC-COVID.
Our ``base'' reranker, called monoT5, is based on T5~\cite{Raffel:1910.10683:2019}.

Given a query $q$ and a set of candidate documents $D$ from Pyserini, for each $d \in D$ we construct the following input sequence to feed into our model:
\begin{equation}
\text{Query: } q \text{ Document: } d \text{ Relevant:}
\end{equation}
\noindent The model is fine-tuned to produce either ``true'' or ``false'' depending on whether the document is relevant or not to the query.
That is, ``true'' and ``false'' are the ground truth predictions in the sequence-to-sequence task, what we call the ``target words''.

At inference time, to compute probabilities for each query--document pair, we apply softmax only to the logits of the ``true'' and ``false'' tokens.
We rerank the candidate documents according to the probabilities assigned to the ``true'' token.
See~\citet{Nogueira_etal_arXiv2020_T5} for additional details about this logit normalization trick and the effects of different target words.

Since in the beginning we did not have training data specific to COVID-19, we fine-tuned our model on the MS MARCO passage dataset~\citep{nguyen2016ms}, which comprises 8.8M passages obtained from the top 10 results retrieved by the Bing search engine (based on around 1M queries).
The training set contains approximately 500k pairs of query and relevant documents, where each query has one relevant passage on average; non-relevant documents for training are also provided as part of the training data.
\citet{Nogueira_etal_arXiv2020_T5} and \citet{Yilmaz_etal_EMNLP2019} have both previously demonstrated that models trained on MS MARCO can be directly applied to other document ranking tasks.

We fine-tuned our monoT5 model with a constant learning rate of $10^{-3}$ for 10k iterations with class-balanced batches of size 128.
We used a maximum of 512 input tokens and one output token (i.e., either ``true'' or ``false'', as described above).
In the MS MARCO passage dataset, none of the inputs required truncation when using this length limit.
Training variants based on T5-base and T5-3B took approximately 4 and 40 hours, respectively, on a single Google TPU v3-8.

At inference time, since output from Pyserini is usually longer than the length restrictions of the model, it is not possible to feed the {\it entire} text into our model at once.
To address this issue, we first segment each document into spans by applying a sliding window of 10 sentences with a stride of~5.
We obtain a probability of relevance for each span by performing inference on it independently, and then select the highest probability among the spans as the relevance score of the document.

\smallskip \noindent {\bf duoT5.} 
A pairwise reranker estimates the probability $s_{i,j}$ that candidate $d_i$ is more relevant than $d_j$ for query $q$, where $i \neq j$.
\citet{nogueira2019multistage} demonstrated that a pairwise BERT reranker running on the output of a pointwise BERT reranker yields statistically significant improvements in ranking metrics.
We applied the same intuition to T5 in a pairwise reranker called duoT5, which takes as input the sequence:
\begin{align*}
\text{Query: } q \text{ Document0: } d_i
\text{ Document1: } d_j \text{ Relevant:}
\end{align*}
where $d_i$ and $d_j$ are unique pairs of candidates from the set $D$.
The model is fine-tuned to predict ``true'' if candidate $d_i$ is more relevant than $d_j$ to query $q$ and ``false'' otherwise. 
We fine-tuned duoT5 using the same hyperparameters as monoT5.

At inference time, we use the top 50 highest scoring documents according to monoT5 as our candidates $\{d_i\}$.
We then obtain probabilities $p_{i,j}$ of $d_i$ being more relevant than $d_j$ for all unique candidate pairs $\{d_i, d_j\}, \forall i \neq j$.
Finally, we compute a single score $s_i$ for candidate $d_i$ as follows:
\begin{equation}
\label{eq:pairwise_sym_sum}
s_i = \sum_{j \in J_i} \left( p_{i,j} + (1-p_{j,i}) \right)
\end{equation}
\noindent where $J_i=\{0 \leq j < 50, j \neq i\}$.
Based on exploratory studies on the MS MARCO passage dataset, this setting leads to the most stable and effective rankings.

\smallskip \noindent {\bf Relevance Feedback.}
The setup of TREC-COVID (see Section~\ref{section:trec-description}) provides a feedback setting where systems can exploit a limited number of relevance judgments on a per-query basis.
How do we take advantage of such training data?
Despite work on fine-tuning transformers in a few-shot setting~\cite{Zhang:2006.05987:2020,LeeCheolhyoung_etal_ICLR2020}, we were wary of the dangers of overfitting on limited data, particularly since there is little guidance on relevance feedback using transformers in the literature.
Instead, we implemented a robust approach that treats relevance feedback as a document classification problem using simple linear classifiers, described in~\citet{Yu_etal_ECIR2019} and \citet{Lin_arXiv2019}.

The approach is conceptually simple:\ for each query, we train a linear classifier (logistic regression) that attempts to distinguish relevant from non-relevant documents {\it for that query}.
The classifier operates on sparse bag-of-words representations using tf--idf term weighting.
At inference time, each candidate document is fed to the classifier, and the classifier score is then linearly interpolated with the original candidate document score to produce a final score.
We describe the input source documents in Section~\ref{section:results}.

\medskip \noindent
All components above have also been open sourced.
The two neural reranking modules are available in PyGaggle,\footnote{\url{pygaggle.ai}} which is our recently developed neural ranking library designed to work with Pyserini.
Our classification-based approach to feedback is implemented in Pyserini directly.
These components are available for integration into any system.

\section{The Covidex}
\label{section:covidex}

Beyond sharing our keyword search infrastructure and reranking models, we've built the Covidex as an operational search engine to demonstrate our capabilities to domain experts who are not interested in individual components.
As deployed, we use the paragraph index and monoT5-base as the reranker.
An additional highlighting module based on BioBERT is described in~\citet{Zhang_etal_arXiv2020_Covidex}.
To decrease end-to-end latency, we rerank only the top 96 documents per query and truncate reranker input to a maximum of 256 tokens.

The Covidex is built using the FastAPI Python web framework, where all incoming API requests are handled by a service that performs searching, reranking, and text highlighting.
Search is performed with Pyserini (Section~\ref{section:keyword}), and the results are then reranked with PyGaggle (Section~\ref{section:reranker}).
The frontend (which is also open source) is built with React to support the use of modular, declarative JavaScript components,\footnote{\url{reactjs.org}} 
taking advantage of its vast ecosystem.

\begin{figure}[t]
\centering
\includegraphics[width=1.0\linewidth]{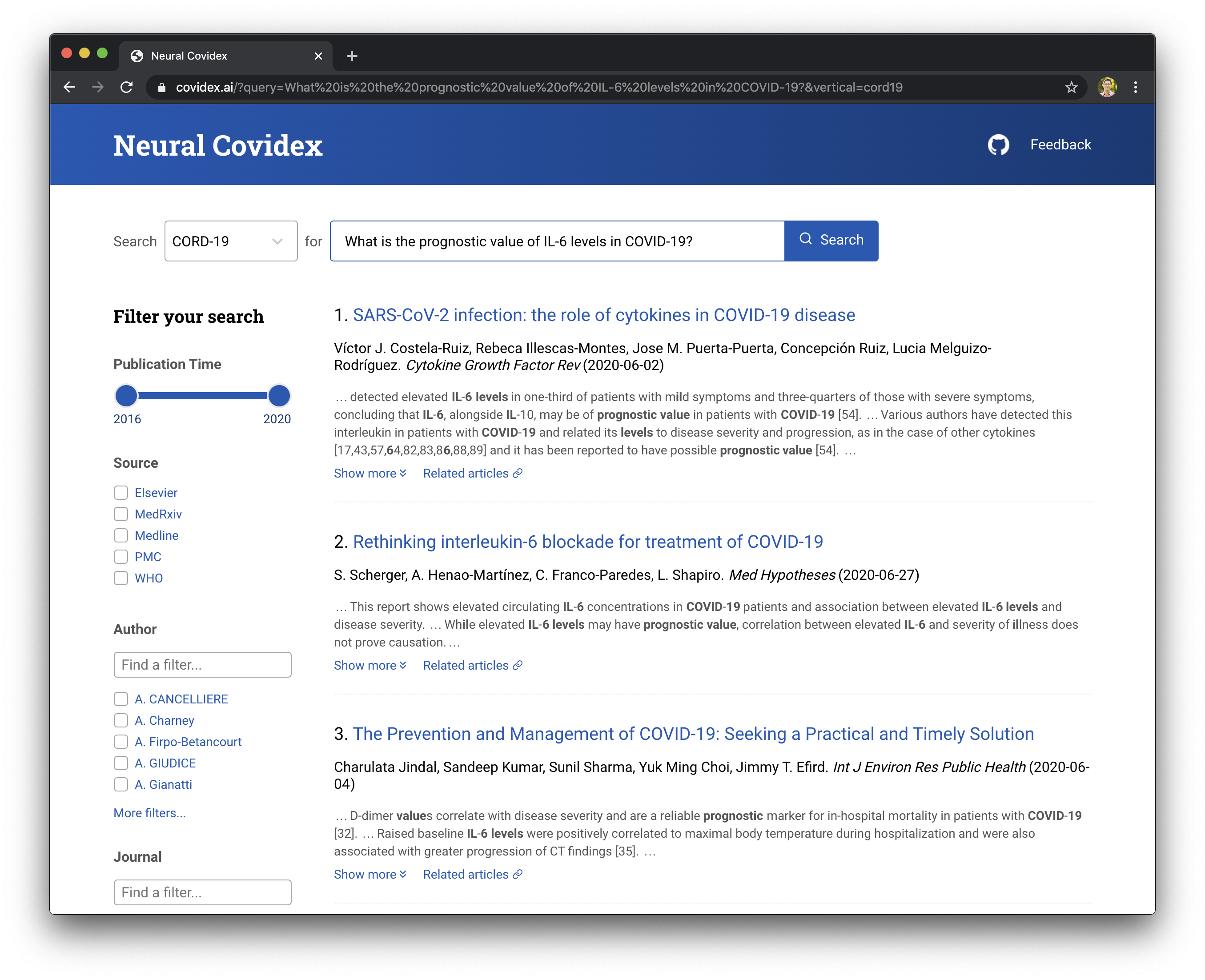}
\caption{Screenshot of the Covidex.}
\label{fig:screenshot}
\end{figure}

A screenshot of our system is shown in Figure~\ref{fig:screenshot}.
Covidex provides standard search capabilities, either based on keyword queries or natural-language input.
Users can click ``Show more'' to reveal the abstract as well as excerpts from the full text, where potentially relevant passages are highlighted.
Clicking on the title brings the user to the article's source on the publisher's site.
In addition, we have implemented a faceted browsing feature.
From CORD-19, we were able to easily expose facets corresponding to dates, authors, journals, and sources.
Navigating by year, for example, allows a user to focus on older coronavirus research (e.g., on SARS) or the latest research on COVID-19, and a combination of the journal and source facets allows a user to differentiate between preprints and the peer-reviewed literature, and between venues with different reputations.

The system is currently deployed across a small cluster of servers, each with two NVIDIA V100 GPUs, as our pipeline requires neural network inference at query time.
Each server runs the complete software stack in a simple replicated setup (no partitioning).
On top of this, we leverage Cloudflare as a simple load balancer, which uses a round robin scheme to dispatch requests across the different servers.
The end-to-end latency for a typical query is around two seconds.

The first implementation of our system was deployed in late March, and we have been incrementally adding features since.
Based on Cloudflare statistics, our site receives around two hundred unique visitors per day and the site serves more than one thousand requests each day.
Of course, usage statistics were (up to several times) higher when we first launched due to publicity on social media.
However, the figures cited above represent a ``steady state'' that has held up over the past few months, in the absence of any deliberate promotion.

\section{TREC-COVID}
\label{section:trec}

Reliable, large-scale evaluations of text retrieval methods are a costly endeavour, typically beyond the resources of individual research groups.
Fortunately, the community-wide TREC-COVID challenge sponsored by the U.S.\ National Institute for Standards and Technology (NIST) provides a forum for evaluating our techniques.

\subsection{Evaluation Overview}
\label{section:trec-description}

The TREC-COVID challenge, which began in mid-April and is still ongoing, provides an opportunity for researchers to study methods for quickly standing up information access systems, both in response to the current pandemic and to prepare for similar future events.

Both out of logistic necessity in evaluation design and because the body of scientific literature is rapidly expanding, TREC-COVID is organized into a series of ``rounds'', each of which use the CORD-19 collection at a snapshot in time.
For a particular round, participating teams develop systems that return results to a number of information needs, called ``topics''---one example is ``serological tests that detect antibodies of COVID-19''.
These results comprise a run or a submission.
NIST then gathers, organizes, and evaluates these runs using a standard pooling methodology~\cite{Voorhees_CLEF2002}.

The product of each round is a collection of relevance judgments, which are annotations by domain experts about the relevance of documents with respect to topics.
On average, there are around 300 judgments (both positive and negative) {\it per topic} from each round.
These relevance judgments are used to evaluate the effectiveness of systems (populating a leaderboard) and can also be used to train machine-learning models in future rounds.
Runs that take advantage of these relevance judgments are known as ``feedback runs'', in contrast to ``automatic'' runs that do not.
A third category, ``manual'' runs, can involve human input, but we did not submit any such runs.

Currently, TREC-COVID has completed round 3 and is in the middle of round 4.
We present evaluation results from rounds 1, 2, and 3, since results from round 4 are not yet available.
Each round contains a number of topics that are persistent (i.e., carryover from previous rounds) as well as new topics.
To avoid retrieving duplicate documents, the evaluation adopts a residual collection methodology, where judged documents (either relevant or not) from previous rounds are automatically removed from consideration.
Thus, for each topic, future rounds only evaluate documents that have not been examined before (either newly published articles or have never been retrieved).
Note that due to the evaluation methodology, scores across rounds {\it are not} comparable.

\begin{table*}[t]
\begin{center}
\begin{small}
\begin{tabular}{lllrrrH}
\toprule
{\bf Team } & {\bf Run} & Type & nDCG@10 & P@5 & mAP & Judged@5 \\
\toprule
\multicolumn{2}{l}{\textbf{Round 1}: 30 topics} \\
sabir & \run{sabir.meta.docs} & automatic & 0.6080 & 0.7800 & 0.3128 & 1.0000\\
GUIR\_S2 & \run{run2}$^{\dagger}$ & automatic & 0.6032 & 0.6867 & 0.2601 & 0.8800 \\
covidex & \run{T5R1} (= monoT5) & automatic & 0.5223 & 0.6467 & 0.2838 & 1.0000\\
\midrule
\multicolumn{2}{l}{\textbf{Round 2}: 35 topics} \\
mpiid5 & \run{mpiid5\_run3}$^{\dagger}$ & manual & 0.6893 & 0.8514 & 0.3380 & 1.0000 \\
CMT & \run{SparseDenseSciBert}$^{\dagger}$ & feedback & 0.6772 & 0.7600 & 0.3115 & 1.0000 \\
GUIR\_S2 & \run{GUIR\_S2\_run1}$^{\dagger}$ & automatic & 0.6251 & 0.7486 & 0.2842 & 0.9885 \\
covidex & \run{covidex.t5} (= monoT5) & automatic & 0.6250 & 0.7314 & 0.2880 & 1.0000 \\
anserini & \run{r2.fusion2} & automatic & 0.5553 & 0.6800 & 0.2725 & 1.0000 \\
anserini & \run{r2.fusion1} & automatic & 0.4827 & 0.6114 & 0.2418 & 1.0000 \\
\midrule
\multicolumn{2}{l}{\textbf{Round 3}: 40 topics} \\
covidex & \run{r3.t5\_lr} & feedback & 0.7740 & 0.8600 & 0.3333 & 0.9700 \\
BioinformaticsUA & \run{BioInfo-run1} & feedback & 0.7715 & 0.8650 & 0.3188 & 0.9750 \\
SFDC & \run{SFDC-fus12-enc23-tf3}$^{\dagger}$ & automatic & 0.6867 & 0.7800 & 0.3160 & 0.9850 \\
covidex & \run{r3.duot5} (= monoT5 + duoT5) & automatic & 0.6626 & 0.7700 & 0.2676 & 0.8850 \\
covidex & \run{r3.monot5} (= monoT5) & automatic & 0.6596 & 0.7800 & 0.2635 & 0.9300 \\
anserini & \run{r3.fusion2} & automatic & 0.6100 & 0.7150 & 0.2641 & 0.9550 \\
anserini & \run{r3.fusion1} & automatic & 0.5359 & 0.6100 & 0.2293 & 0.8950 \\
\bottomrule
\end{tabular}
\end{small}
\end{center}
\caption{Selected TREC-COVID results. Our submissions are under teams ``covidex'' and ``anserini''. All runs notated with $^{\dagger}$ incorporate our infrastructure components in some way.}
\label{tab:results}
\end{table*}

\subsection{Results}
\label{section:results}

A selection of results from TREC-COVID are shown in Table~\ref{tab:results}, where we report standard metrics computed by NIST.
We submitted runs under team ``covidex'' (for neural models) and team ``anserini'' (for our bag-of-words baselines).

\smallskip
\noindent In {\bf Round 1}, there were 143 runs from 56 teams.
Our best run \run{T5R1} used BM25 for first-stage retrieval using the paragraph index followed by our monoT5-3B reranker, trained on MS MARCO (as described in Section~\ref{section:reranker}).
The best automatic neural run was \run{run2} from team GUIR\_S2~\cite{MacAvaney:2005.02365:2020}, which was built on Anserini.
This run placed second behind the best automatic run, \run{sabir.meta.docs}, which interestingly was based on the vector-space model.

While we did make meaningful infrastructure contributions (e.g., Anserini provided the keyword search results that fed the neural ranking models of team GUIR\_S2), our own run \run{T5R1} was substantially behind the top-scoring runs.
A post-hoc experiment with round 1 relevance judgments showed that using the paragraph index did not turn out to be the best choice:\
simply replacing with the abstract index (but retaining the monoT5-3B reranker) improved nDCG@10 from 0.5223 to 0.5702.\footnote{Despite this finding, we suspect that there may be evaluation artifacts at play here, because our impressions from the deployed system suggest that results from the paragraph index are better. Thus, the deployed Covidex still uses paragraph indexes.}

We learned two important lessons from the results of round 1:

\begin{enumerate}[leftmargin=*]
\itemsep0em 

\item The effectiveness of simple rank fusion techniques that can exploit diverse ranking signals by combining multiple ranked lists.
Many teams adopted such techniques (including the top-scoring run), which proved both robust and effective.
This is not a new observation in information retrieval, but is once again affirmed by TREC-COVID.

\item
The importance of building the ``right'' query representations for keyword search.
Each TREC-COVID topic contains three fields:\ query, question, and narrative.
The query field describes the information need using a few keywords, similar to what a user would type into a web search engine.
The question field phrases the information need as a well-formed natural language question, and the narrative field contains additional details in a short paragraph.
The query field may be missing important keywords, but the other two fields often contain too many ``noisy'' terms unrelated to the information need.

Thus, it makes sense to leverage information from multiple fields in constructing keyword queries, but to do so selectively.
Based on results from round 1, the following query generation technique proved to be effective: 
when constructing the keyword query for a given topic, we take the non-stopwords from the query field and further expand them with terms belonging to named entities extracted from the question field using ScispaCy~\cite{Neumann:2019}. 

\end{enumerate}

\noindent We saw these two lessons as an opportunity to further contribute community infrastructure, and starting in round 2 we made two fusion runs from Anserini freely available: \run{fusion1} and \run{fusion2}.
In both runs, we combined rankings from the abstract, full-text, and paragraph indexes via reciprocal rank fusion (RRF)~\cite{Cormack_etal_SIGIR2009}.
The runs differed in their treatment of the query representation.
The run \run{fusion1} simply took the query field from the topics as the basis for keyword search, while run \run{fusion2} incorporated the query generator described above to augment the query representation with key phrases.
These runs were made available {\it before} the deadline so that other teams could use them, and indeed many took advantage of them.

\smallskip
\noindent In {\bf Round 2}, there were 136 runs from 51 teams.
Our two Anserini baseline fusion runs are shown as \run{r2.fusion1} and \run{r2.fusion2} in Table~\ref{tab:results}.
Comparing these two fusion baselines, we see that our query generation approach yields a large gain in effectiveness.
Ablation studies further confirmed that ranking signals from the different indexes do contribute to the overall higher effectiveness of the rank fusion runs.
That is, the effectiveness of the fusion results is higher than results from any of the individual indexes.

Our \run{covidex.t5} run takes \run{r2.fusion1} and \run{r2.fusion2}, reranks both with monoT5-3B, and then combines (with RRF) the outputs of both.
The monoT5-3B model was fine-tuned on MS MARCO then fine-tuned (again) on a medical subset of MS MARCO~\cite{MacAvaney:2005.02365:2020}.
This run essentially tied for the best automatic run \run{GUIR\_S2\_run1}, which scored just $0.0001$ higher.

As additional context, Table~\ref{tab:results} shows the best ``manual'' and ``automatic'' runs from round 2 (\run{mpiid5\_run3} and \run{SparseDenseSciBert}, respectively), which were also the top two runs overall.
These results show that manual and feedback techniques can achieve quite a bit of gain over fully automatic techniques.
Both of these runs and four out of the five top teams in round 2 took advantage of the fusion baselines we provided, which demonstrates our impact not only in developing effective ranking models, but also our service to the community in providing infrastructure.

\smallskip
\noindent In {\bf Round 3}, there were 79 runs from 31 teams.
Our Anserini fusion baselines, \run{r3.fusion1} and \run{r3.fusion2}, remained the same from the previous round and continued to provide strong baselines.

Our run \run{r3.duot5} represents the first deployment of our monoT5 and duoT5 multi-stage reranking pipeline (see Section~\ref{section:reranker}), which is a fusion of the fusion runs as the first-stage candidates, reranked by monoT5 and then duoT5.
From Table~\ref{tab:results}, we see that duoT5 does indeed improve over just using monoT5 (run \run{r3.monot5}), albeit the gains are small (but we found that the duoT5 run has more unjudged documents).
The \run{r3.duot5} run ranks second among all teams under the ``automatic'' condition, and we are about two points behind team SFDC.
However, according to~\citet{Esteva:2006.09595:2020}, their general approach incorporates Anserini fusion runs, which bolsters our case that we are providing valuable infrastructure for the community.

Our own feedback run \run{r3.t5\_lr} implements the classification-based feedback technique (see Section~\ref{section:reranker}) with monoT5 results as the input source document (with a mixing weight of 0.5 to combine monoT5 scores with classifier scores).
This was the highest-scoring run across all submissions (all categories), just a bit ahead of \run{BioInfo-run1}.

\section{Conclusions}

Our project has three goals:\ build community infrastructure, advance the state of the art in neural ranking, and provide a useful application.
We believe that our efforts can contribute to the fight against this global pandemic.
Beyond COVID-19, the capabilities we've developed can be applied to analyzing the scientific literature more broadly.

\section{Acknowledgments}

This research was supported in part by the Canada First Research Excellence Fund, the Natural Sciences and Engineering Research Council (NSERC) of Canada, CIFAR AI \& COVID-19 Catalyst Funding 2019--2020, and Microsoft AI for Good COVID-19 Grant.
We'd like to thank Kyle Lo from AI2 for helpful discussions and Colin Raffel from Google for his assistance with T5.

\bibliographystyle{acl_natbib}
\bibliography{main}

\end{document}